\documentclass{elsart}

\usepackage{color}

\usepackage{rbnca}
\usepackage[dvips]{graphicx}
\usepackage{cite}
\graphicspath{{./images/},{./figs/}}

\usepackage{algorithmic}
\usepackage{algorithm}

\usepackage{amssymb}

\newcommand{\bfs}{\bsym{s}}
\newcommand{\bfS}{\bsym{S}}

\DeclareSymbolFont{AMSb}{U}{msb}{m}{n}
\DeclareMathSymbol{\R}{\mathbin}{AMSb}{"52}

\begin{document}

\begin{frontmatter}
  
\title{Deducing Local Rules for Solving Global Tasks with Random
  Boolean Networks}

\author[aut1]{Bertrand Mesot} 
\ead{bertrand.mesot@idiap.ch}
\ead[url]{http://www.idiap.ch/$\sim$bmesot}
\author[aut2]{Christof Teuscher\thanksref{lanl}} 
\ead{christof@teuscher.ch}
\ead[url]{http://www.teuscher.ch/christof}

\address[aut1]{IDIAP Research Institute, Rue de Simplon 4, CH--1920,
  Martigny, Switzerland} 
\address[aut2]{Los Alamos National Laboratory, Advanced Computing
  Laboratory, CCS-1, MS-B287, Los Alamos, NM 87545, USA} 

\thanks[lanl]{The work in this paper was completed while affiliated
    with the University of California, San Diego (UCSD), and supported
    by grant PBEL2--104420 from the Swiss National Science
    Foundation.}

\begin{abstract}
It has been shown that uniform as well as non-uniform cellular
automata (CA) can be evolved to perform certain computational
tasks. Random Boolean networks are a generalization of two-state
cellular automata, where the interconnection topology and the cell's
rules are specified at random.
    
Here we present a novel analytical approach to find the local rules of
random Boolean networks (RBNs) to solve the global density
classification and the synchronization task from any initial
configuration.  We quantitatively and qualitatively compare our
results with previously published work on cellular automata and show
that randomly interconnected automata are computationally more
efficient in solving these two global tasks. Our approach also
provides convergence and quality estimates and allows the networks to
be randomly rewired during operation, without affecting the global
performance. Finally, we show that RBNs outperform small-world
topologies on the density classification task and that they perform
equally well on the synchronization task.
  
Our novel approach and the results may have applications in designing
robust complex networks and locally interacting distributed computing
systems for solving global tasks.
\end{abstract}

\end{frontmatter}

\section{Introduction}
\label{sec:intro}
{\sl Cellular automata} (CA) \cite{wolfram84,toffoli87} were
originally conceived by Ulam and von Neumann \cite{neumann66:_theor}
in the 1940s to provide a formal framework for investigating the
behavior of complex, extended systems. CAs are dynamical systems in
which space and time are discrete. A cellular automaton usually
consists of a $D$-dimensional regular lattice of $N$ lattice sites,
commonly called {\sl cells}. Each cell can be in one of a finite
number of $S$ possible states and further consists of a transition
function $F$ (also called {\sl rule}), which maps the neighboring
states to the set of cell states. CAs are called {\sl uniform} if all
cells contain the same rule, otherwise they are {\sl non-uniform}.
Each cell takes as input the states of the cells within some finite
local neighborhood.  By convention, a cell is considered to be a
member of its own neighborhood. In the standard one-dimensional CA
model, for example, each cell is connected to the $r$ ($r$ stands for
radius) immediate local neighbors on either side as to itself, thus
each cell is connected to $2r+1$ neighbors.

{\sl Random Boolean networks} (RBNs), on the other hand, form a more
general class of discrete dynamical systems, in which two-state
cellular automata are a special subclass. Random networks and RBNs
were investigated in the past forty years by many a researcher (see
for example \cite{asbhy1966,rozonoer69,allanson56,amari71}), but
became popular mainly due to the contributions of Stuart Kauffman
\cite{kauffman69,kauffman84,kauffman93} and others.

In its simplest form, a RBN is composed of $N$ nodes (sometimes also
called {\sl elements} or {\sl cells}) where each node can be in one of
two possible states ($0$ or $1$) and receives inputs from $K$ randomly
chosen other nodes (self-connections are allowed). Note that $K$ can
refer to the {\sl exact} or to the {\sl average} number of connections
between the nodes. In the geneticist's term, for example, $K$ measures
the richness of epistatic interactions among the components of a
system \cite{kauffman93}.

The deterministic behavior of each node is specified by one out of the
$2^{2^K}$ Boolean functions, specifying its next value for each of the
$2^K$ combinations with $K$ inputs. Like CAs, RBNs can be {\sl
non-uniform} (i.e., each node can potentially have a different rule)
or {\sl uniform}, although they are non-uniform in the majority of the
cases.  The network's nodes are usually updated synchronously,
although multiple asynchronous updating schemes exist (see
\cite{gershenson2003:alife} for an overview).

\begin{figure}[htb]
  \centering \includegraphics[width=.7\textwidth]{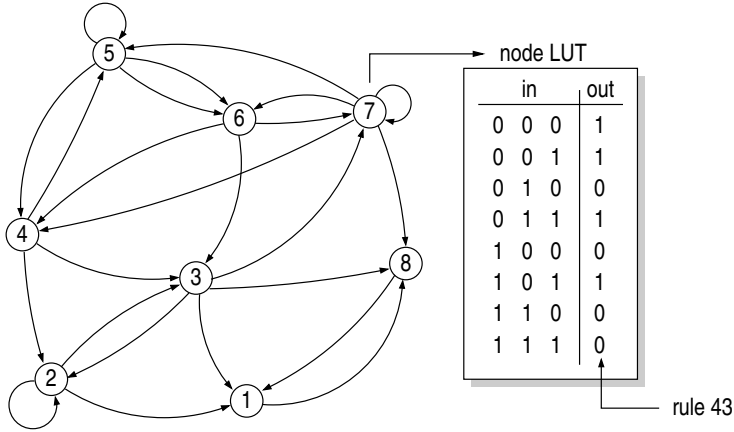}
  \caption{Illustration of a random Boolean network with $N=8$ nodes
    and $K=3$ inputs per node (self-connections are allowed).  The
    node rules are commonly represented by {\sl lookup-tables} (LUTs),
    which associate a $1$-bit output (the node's future state) to each
    possible $K$-bit input configuration. The table's out-column is
    commonly called the {\sl rule} of the node.}
  \label{fig:rbn}
\end{figure}

Synchronous random Boolean networks as introduced by Kauffman are
commonly called {\sl NK} networks or models. Figure \ref{fig:rbn}
shows a possible {\sl NK} random Boolean network representation ($N=8,
K=3$). The model is very similar to the well-studied class of models
which arises in statistical physics, called {\sl spin-glasses}
\cite{edwards75}.  Spin-glasses are disordered magnetic materials in
which the orientation of nearby magnetic dipoles may be either
parallel or antiparallel in space.

More formally speaking, let $\bfs^t$ be the state-vector of the
network's nodes at time $t$. The neighborhood $\Pi(x_i)$ of a node
$x_i$ is defined as
\[\Pi(x_i) \subset P_K(\{x_1, \dots, x_N\}),\]
where $P_K(X)$ denotes the set of all subsets of size $K$ from $X$
(self-connections are allowed). The neighborhood size is given by
$K=|\Pi(x_i)|$. The most common ways to choose the neighborhood are
the following:
\begin{itemize}
\item {\sl random neighborhood}: choose the $K$ neighbors randomly
  from the set \\$\{x_1, \dots, x_N\} \setminus \{x_i\}$ (without
  self-connections) or from the set $\{x_1, \dots, x_N\}$
  (self-connections allowed)
  
\item {\sl adjacent neighborhood}: $K$ neighbors are
  randomly chosen in the ``immediate'' neighborhood.
\end{itemize}
$A(N,K)$ is sometimes used to denote a {\sl NK} model with adjacent
neighborhood, $N(N,K)$ for the random neighborhood. We shall not go
into more details of the adjacent neighborhood here, as we will only
consider purely random networks with a uniform probability to connect
two nodes together in this article.

Figure \ref{fig:rbn} shows the interaction graph $G(V,E)$ where
$V=\{x_1,\dots, x_N\}$ corresponds to the set of nodes, and
$\{x_i,x_j\} \in E$ if and only if $x_i$ is connected to $x_j$. The
node rules are commonly represented by {\sl lookup-tables} (LUTs),
which associate a $1$-bit output (the node's future state) to each
possible $K$-bit input configuration. The table's out-column is
commonly called the {\sl rule} of the node.

Whereas the homogeneous connectivity and the usually uniform rules of
a CA restrict the automaton's parameter space, RBNs have a vastly
greater space, which adds an additional challenge for both
evolutionary and analytical methods, either to analyze the networks'
behavior or to find networks able to perform a given task.

The main contribution of this paper consists in a novel analytical
method which allows to determine the local rules of a random Boolean
network for two global and well-known problems for cellular automata,
(1) the synchronization and (2) the density classification
task. Whereas previous work was purely quantitative and focused on the
co-evolution of cellular automata to find suitable rules and
architectures \cite{sipper97phyd}, our qualitative approach goes a
step further and also provides convergence and quality estimates. We
systematically compare the results with previous work on standard and
non-standard cellular automata architectures, including small-world
networks, and show that randomly interconnected automata perform
better than a standard cellular automata with a regular
interconnection architecture. Our method also allows the networks to
be randomly rewired during operation without affecting the global
performance. The approach and results may have applications in
designing robust complex networks and locally interacting distributed
computing systems for solving global tasks.

In Section \ref{sec:tasks} we describe the two global tasks used in
this paper, the density classification task and the synchronization
task. The main challenge in solving such global tasks with CAs and
RBNs consists in finding the node's local rules that will result in
the desired global automata behavior. Note that no generally
applicable method for this exists. In Section \ref{sec:comp_and_inf}
we derive the recursive equation which gives the probability of a node
to be in state $1$ at the next time step. This equation is then used
in Section \ref{sec:rules} to determine the node's rules for both
tasks. Section \ref{sec:experiments} examines the network's
performance as a function of $N$ and $K$ whereas Section
\ref{sec:rbns_vs_cas} compares the results with previously published
performance results of cellular automata for the same tasks. In order
to measure the rule's capacity to solve the tasks, we introduce an
entropy-based performance measure in Section
\ref{sec:entropy_measure}. The same measure also allows us to predict
how quickly the two tasks can be solved. In Section
\ref{sec:non-standard} we briefly discuss non-uniform random Boolean
networks and perform experiments with small-world network topologies.
Our findings, their possible future applications, and future work are
discussed in Section \ref{sec:conclusions}.

\section{Task Definitions}
\label{sec:tasks}
Two commonly used applications for CAs are the {\sl density} and the
{\sl synchronization} task. Both of these ``global'' tasks are mostly
trivial to solve if one has a global view on the system (i.e., if one
has access to the state of all nodes at the same time), but are
non-trivial to solve for CA and RBNs, mainly because of the locality
and the limited number of their interconnections, the simplicity of
the basic cells (i.e., the basic components), and because there is no
generally valid way to determine the cells's rules for such a massive
parallel system. Unlike in the standard approach to parallel
computation, in which a given problem is split into independent
sub-problems, CAs and RBNs have to solve problems by the bottom-up
approach, where the global and usually complex behavior arises from
nonlinear, spatially extended, and local interactions
\cite{mitchell94}. This difficulty has naturally also limited CAs
applications.

In the following, the density classification and the synchronization
task shall be briefly described.

\subsection{The Density Classification Task}
\label{sec:density_task}
The density classification task for CAs must decide whether or not the
initial configuration of the automaton contains more than $50\%$ of
$1$s. In this context, the term ``configuration'' refers to an
assignment of the states $0$ or $1$ to each cell of the CA (i.e.,
there are $2^N$ possible initial configurations). The desired behavior
of the automaton is to have all of its cells set to 1 if the initial
density of $1$s exceeded $1/2$, and all $0$ otherwise. The density
task for CAs can straightforwardly be applied to RBNs. The special
case of having an equal number of $1$s and $0$s in the network is
commonly avoided by using an odd number of nodes.

The density task was studied among others by Mitchell et al.
\cite{mitchell94,mitchelletal93,mitchelletal96}, Das et al.
\cite{dasetal94,dasetal95}, Sipper
\cite{sipper95:alife,sipper95:ecal,sipper94:alife,sipper96:physicad},
Sipper et al. \cite{sipper99:tcs,sipper97phyd,sipper98:ijmpc},
Capcarr\`ere \cite{capcarrere02}, Capcarr\`ere et al.
\cite{capcarrere96,capcarrere01,capcarrere99} and Tomassini et al.
\cite{tomassini02:cs} in various forms, including non-uniform CAs,
asynchronous CAs, and non-standard architectures. It should also be
mentioned that no uniform two-state CA exists, which perfectly solves
the density classification task for all initial configurations
\cite{land95}.

Let $n_1^t$ be the number of nodes being in state $1$ at time $t$. The
density classification task can then be formally described by the
following equation:
\beq
\exists t_{\rm lim}\quad\textrm{s.t.}\quad t \geqslant t_{\rm%
  lim}\quad\Rightarrow\quad n_1^t = \left\{%
\begin{array}{cl}
0           & \textrm{if $n_1^0 < N/2$}\\
\noalign{\vspace{5pt}}
N           & \textrm{if $n_1^0 > N/2$}\\
\noalign{\vspace{5pt}}
{\displaystyle\frac{N}{2}} & \textrm{otherwise}
\end{array}\right..
\label{eq:density}
\eeq
Let us define $p_1^t = n_1^t/N$ as the fraction of nodes that are in
state $1$ at time $t$. We will sometimes use $p_1^t$ instead of
$n_1^t$ when the networks are sufficiently large (i.e., $N \to
+\infty$), since $p_1^t$ may then be interpreted as the probability of
the node to be in state $1$ at time $t$. Moreover, since we only
consider a synchronous updating scheme, $p_1^{t+1}$ will solely depend
on $p_1^t$. The network's behavior can thus simply be described by a
map of the form $p_1^{t+1} = Q(p_1^t)$, where $Q:\R \longmapsto
\R$. Each of the three solutions of the density classification task
will then be represented by a fixed point in this map. Figure
\ref{fig:density_poly} shows three possible functions $Q$ that satisfy
Equation \ref{eq:density}. Note that the fixed point in $p_1^* = 1/2$
is unstable, which means that a small perturbation of the network's
state, such as for example one state change among the $N$ nodes, will
drive the system towards the wrong fixed point, i.e., towards the
wrong solution. This unstable fixed point can fortunately be avoided
by using an odd number of nodes.

\iclfig{scale=0.8}{density_poly.0}{Three possible maps which satisfy
Equation \ref{eq:density}. The dash-dotted curve represents the one
that will take the highest number of iterations before settling down
in a fixed point.}  {fig:density_poly}

\subsection{The Synchronization Task}
\label{sec:sync_task}
The one-dimensional {\sl synchronization task} (also called {\sl
  firefly task}) for synchronous CA was introduced by Das et al.
\cite{dasetal95} and studied among others by Hordijk \cite{hordijk96}
and Sipper \cite{sipper97:book}. In this task, the two-state
one-dimensional or two-dimensional CA, given any initial
configuration, must reach a final configuration within $M$ time steps,
that oscillates between all $0$s and all $1$s on successive time
steps. The whole automaton is then globally synchronized.

As with the density classification task, synchronization is a
non-trivial global computation task for a small-radius CA where all
cells must coordinate their behavior with all the other cells while
having only a very local view of their neighborhood. The two-state
machine of each cell obviously also prevents any counting, which would
make the task much less difficult.  Furthermore, in the non-uniform
case, where each cell can have a different rule, there is an immediate
solution consisting of a unique ``master'' rule that alternates
between 0 and 1 without looking at his neighbors, and all other cells
being its ``slave'', i.e., alternating according to their right (or
upper, in the two-dimensional case) neighbor state only. However,
Sipper \cite{sipper97:book} used non-uniform CAs to find perfect
synchronizing CAs by means of evolution only. It appeared that this
very particular solution was never found by evolution and, in fact,
the ``master'' or ``blind'' rule {\tt 10101010} (rule 170) in one
dimension, was never part of the evolved solutions. This is simply due
to the fact that this rule has to be unique for the solution to be
perfect, which is contradictory to the natural tendency of the
evolutionary algorithm used, as was demonstrated by Capcarr\`ere
\cite{capcarrere99}.

The two-dimensional version of the task is identical to the
one-dimensional version, except that the necessary number of time
steps granted to synchronize is not anymore in the order of $N$, the
number of cells in the automaton, but in the order of $n+m$, where $n$
and $m$ are the size of each side of the standard CA. Therefore, the
speed of synchronization is much faster compared to the
one-dimensional version.

Extending the two-dimensional synchronization task for CAs to RBNs is
again straightforward and solutions found be means of co-evolutionary
algorithms were presented by Teuscher and Capcarr\`ere in
\cite{teuscher03:ices03}. The firefly synchronization task was also
successfully implemented in actual evolutionary hardware, for both the
one \cite{sipper97:firefly} and the two-dimensional
\cite{teuscher03:ices03} version.

Similarly to the density classification task, the synchronization task
may also be formally described by a condition on the number of $1$s in
the network's state:
\beq
\exists t_{\rm lim}\quad\textrm{s.t.}\quad t \geqslant t_{\rm
  lim}\quad\Rightarrow\quad n_1^t = \left\{%
\begin{array}{cl}
N           & \textrm{if $n_1^{t-1} = 0$}\\
\noalign{\vspace{5pt}}
0           & \textrm{if $n_1^{t-1} = N$}\\
\end{array}\right.
\label{eq:sync}
\eeq
Figure \ref{fig:firefly_poly} shows three possible maps which satisfy
Equation \ref{eq:sync}. Contrary to the density classification task,
fixed points must be avoided since oscillations are required.

\iclfig{scale=0.8}{firefly_poly.0}{Three possible maps which satisfy
  Equation \ref{eq:sync}. The dash-dotted curve requires the highest
  number of iterations before oscillating between all $p_1 = 0$ and
  $p_1 = 1$.}  {fig:firefly_poly}

\section{From Global Configurations to Local Interactions}
\label{sec:comp_and_inf}
The challenge in solving the density classification and
synchronization task with CAs and RBNs consists in finding the node's
local rules that will result in a global automaton behavior which
satisfies Equations \ref{eq:density} or \ref{eq:sync} for all possible
initial configurations. The difficulty comes from the fact that each
node has only access to $K$ randomly chosen bits of the entire RBN's
state vector $\bfs^t$ at time $t$. It is therefore crucial to know how
the global network state is seen by the local nodes in order to
understand the network's dynamics.

Let $W_1^t$ be a random variable representing the number of $1$s in a
node's neighborhood at time $t$ and let $N_1^t$ and $P_1^t = N_1^t/N$
be two random variables which represent the number of $1$s and the
percentage of $1$s in the network's state $\bfs^t$ at time $t$
respectively. $W_1^t$ obviously depends on $P_1^t$, thus, if we
suppose that the nodes' neighbors are randomly chosen from a uniform
distribution, then the probability of having $d$ neighbors in state
$1$ at time $t$, knowing the percentage of $1$s (i.e., $p_1^t$) in the
global network state $\bfs^t$ of a RBN, is therefore given by:
\beq
P(W_1^t = d \,|\, P_1^t = p_1^t) = {K \choose d} (p_1^t)^d\,
(1-p_1^t)^{K-d}.
\label{eq:bin}
\eeq
The distribution of $W_1^t$ knowing $P_1$ thus follows a binomial law
$\mathcal{B}(K, p_1^t)$ and the expected number of $1$s in a node's
neighborhood is therefore $K p_1^t$, which implies that the average
number of $1$s in the input of a node is the same as the average
number of $1$s in the global network state. This property is actually
the key that will allow us to reduce the complex behavior of $N$ nodes
to the stochastic behavior of a single node which behaves on the
average like the entire network.

Let $\bfs^t$ be the state of a RBN at time $t$, composed of $N$ nodes,
each being randomly connected to $K$ other nodes (self-connections are
allowed).  Furthermore, let us assume that $N$ is sufficiently large
to consider $p_1^t = n_1^t/N$ as the probability of a node to be in
state 1 at time $t$. We will later see in our experiments that this
approximation only affects performance. Knowing $\bfs^t$, the
probability that the future state of node $i$ is $1$ is then given by:
\beq
P(S_i^{t+1} = 1 \,|\, \bfS^t = \bfs^t) = \sum_{j=0}^{2^K-1}
P(S_i^{t+1} = 1 \,|\, \Phi_i^t = j) \, P(\Phi_i^t = j \,|\, \bfS^t =
\bfs^t),
\eeq
where $S_i^{t+1}$ and $\Phi_i^t$ are two random variables that
represent the state of node $i$ at time $t+1$ and its input at time
$t$ respectively. This is a probabilistic description of the node's
Boolean transfer function (i.e., its rule), where $P(\Phi_i^t \cd
\bfS^t)$ and $P(S_i^{t+1} \cd \Phi_i^t)$ are the input and output
distributions respectively. If $f_i(j)$ is the output of node $i$ for
an input value $j$, then the previous equation becomes:
\beq
P(S_i^{t+1} = 1 \,|\, \bfS^t = \bfs^t) = \sum_{j=0}^{2^K-1} f_i(j)
\, P(\Phi_i^t = j \,|\, \bfS^t = \bfs^t),
\eeq
since the probability of a node to be in state $1$ at time $t+1$ is
$1$ if $f_i(j) = 1$, and $0$ otherwise. By using the variables $W_1^t$
and $P_1^t$ as introduced above, this latter equation can further be
reduced to:
\beq
P(S_i^{t+1} = 1 \,|\, P_1^t = p_1^t) = \sum_{j=0}^{K} c_j\, P(W_1^t =
j \,|\, P_1^t = p_1^t),
\label{eq:input}
\eeq
where $j$ is the number of $1$s in the input of node $i$ and $c_j$ is
the probability that the future state (the output) of node $i$ is $1$,
provided that the input contains $j$ times a 1.

Putting Equation \ref{eq:bin} into Equation \ref{eq:input} gives us:
\beq
P(S_i^{t+1} = 1 \,|\, P_1^t = p_1^t) = \sum_{j=0}^{K} c_j {K \choose
  j} \left(p_1^t\right)^j \left(1-p_1^t\right)^{K-j},
\eeq
which holds for any node $i$. Since $P(S_i^{t+1} = 1\,|\, P_1^t =
p_1^t)$ is equivalent to $p_1^{t+1}$, the probability for a node to be
in state $1$ at time $t+1$, we can now express $p_1^{t+1}$ as a
function of $p_1^t$ by means of the following recursive equation:
\beq
p_1^{t+1} = \sum_{j=0}^K a_j \left(p_1^t\right)^j
\left(1-p_1^t\right)^{K-j}\quad \textrm{with}\quad a_j = {K \choose j}
c_j\in\left[0,{K \choose j}\right],
\label{eq:poly}
\eeq
where the parameters $a_j$ represent the number of the rule's inputs
that contain $j$ times a $1$ and that produce a $1$ as output. Note
that, if we represent a node's rule by a lookup-table, then $a_j$
corresponds to the number of entries that contain $j$ times a $1$ and
that have a $1$ as output. For example, the two-bit {\sl XOR} rule
(i.e., the output is $1$ if and only if the two inputs are different)
corresponds to $a_0 = 0$, $a_1 = 2$ and $a_2 = 0$ since the output is
$1$ if there is only one $1$ in the input. On the other hand, the
two-bit {\sl NAND} rule corresponds to: $a_0 = 1$, $a_1 = 2$ and $a_2
= 0$.

Recently, Matache and Heidel \cite{matache04} used a formula to
describe the probability of finding a node in state $1$ at time $t$ to
analyze deterministic chaos in random Boolean networks. The basic idea
is similar to our approach, but their work, which extends the model
studied by Andrecut and Ali \cite{andrecut01}, is limited to rule
$126$ only, whereas our approach is valid for any rules.

\section{Finding the Node's Rules}
\label{sec:rules}

\subsection{Rules for the Density Classification Task}
\label{sec:density_rules}
As seen in Section \ref{sec:density_task}, solving the density task
can be reduced to finding maps $p_1^{t+1} = Q_K(p_1^t)$ which satisfy
the constraints that were informally represented in Figure
\ref{fig:density_poly}. More formally speaking, these constraints may
be defined by the following set of equations:
\[{\rm(C1)}\qquad Q_K(0) = 0,\quad Q_K(1) = 1,\quad Q_K(1/2)=1/2,\quad
Q_K''(1/2) = 0\] 
\[{\rm(C2)}\qquad Q_K''(x) > 0\quad \forall x \in [0,1/2[\quad {\rm
      and}\quad Q_K''(x) < 0\quad \forall x \in \;]1/2,1].\] 
However, according to Equation \ref{eq:poly}, we have:
\beq
Q_K(x) = \sum_{j=0}^K a_j\,x^j\,(1-x)^{K-j}\quad \textrm{with}\quad
a_j \in \left[0,{K \choose j}\right].
\label{eq:poly2}
\eeq
Applying constraints C1 and C2 to Equation \ref{eq:poly2} and solving
it as a function of $a_j$ provides us now with an elegant means to
directly determine the node's rules for the density task. In the
following sections, different possible solutions for small
connectivity parameters $K$ shall be presented. We will see that the
simple {\sl majority rule} presents a common solution for all networks
with a connectivity $K>2$.

\subsubsection{$K=2$: No Solution}
For $K=2$, Equation \ref{eq:poly2} becomes:
\[Q_2(x) = a_0 (1-x)^2 + a_1 x (1-x) + a_2 x^2.\]
After applying constraints C1 we obtain the following set of values:
\begin{eqnarray*}
Q_2(0) = 0 & \quad\Rightarrow\quad & a_0 = 0 \\
Q_2(1) = 1 & \quad\Rightarrow\quad & a_2 = 1 \\
Q_2(1/2) = \frac{1}{2^2}(a_1+1) = \frac{1}{2} & \Rightarrow & a_1 = 1
\end{eqnarray*}
This implies $Q_2(x) = x$ and therefore $Q_2''(x) = 0$ for all $x$,
which does not satisfy constraints C2. Hence, for $K=2$ no set of
rules is able to perfectly solve the density classification task for
all initial configurations.

\subsubsection{$K=3$: A First Solution}
\label{sec:k3}
For $K=3$, Equation \ref{eq:poly2} becomes:
\[Q_3(x) = a_0 (1-x)^3 + a_1 x (1-x)^2 + a_2 x^2 (1-x) + a_3 x^3.\]
In order to satisfy constraints C1, we must impose
$a_0 = 0$, $a_3 = 1$ and $a_2 = 3-a_1$. $Q_3(x)$ and $Q_3''(x)$ then
become:
\begin{eqnarray*}
Q_3(x)   & = & 2 (a_1-1) x^3 + 3 (1-a_1) x^2 + a_1 x\\
Q_3''(x) & = & 12 (a_1-1) x + 6 (1-a_1).
\end{eqnarray*}
After applying constraints C2, we obtain:
\[Q_3''(x) > 0\quad\Rightarrow\quad 12 (a_1-1) x > 6
(a_1-1)\quad\Rightarrow\quad x < \frac{1}{2}\quad\textrm{iif}\quad a_1
= 0.\] 
One can easily verify that $a_1 = 0$ is the sole solution. The
corresponding polynomial is therefore:
\beq
Q_3(x) = 3 x^2 (1-x) + x^3.
\label{eq:density_k3}
\eeq
This solution simply corresponds to the well-known {\sl majority}
rule: its output is $1$ if and only if there are more $1$s than $0$s
in the inputs, and $0$ otherwise. As one can verify, since $a_0 = a_1
= 0$, the only input with all bits set to $0$ and the ${K \choose
  1}=3$ inputs with one bit set to $1$ result in a $0$ as output.
Conversely, for all inputs containing more than one bit set to $1$,
the output becomes $1$, since $a_2 = 3$ and $a_3 = 1$.

\subsubsection{$K=4$: Many Rules}
For $K=4$ only one polynomial satisfies both sets of constraints. It
corresponds to: $a_0=0$, $a_1=0$, $a_2=3$, $a_3=4$ and $a_4=1$. Since
$a_2=3$ and according to Equation \ref{eq:poly2}, of the ${K \choose
2} = 6$ inputs with two bits set to $1$, only half of them will set
the output to $1$.  Since there are $20$ possibilities of setting $3$
out of $6$ outputs to $1$, there are $20$ (out of $2^{16}=65536$)
equivalent rules which will successfully solve the density
classification task. These rules are represented in Table
\ref{tab:k4rules}. As one can see, all rules are in a certain sense
{\sl majority} rules, but applied to an even number of connections,
reason why many possibilities exist.
\begin{table}
\begin{center}
\begin{tabular}{|l|l||l|l|}
\hline
Binary & Hex & Binary & Hex \\
\hline
 \tt 1110100011101000 & \tt E8E8 & \tt 1111100011100000 & \tt F8E0 \\
 \tt 1110101010101000 & \tt EAE8 & \tt 1110111011000000 & \tt EEC0 \\
 \tt 1110110010101000 & \tt ECA8 & \tt 1111101011000000 & \tt FAC0 \\ 
 \tt 1111100010101000 & \tt F8A8 & \tt 1111111010000000 & \tt FE80 \\
 \tt 1110100011001000 & \tt E8C8 & \tt 1111110011000000 & \tt FCC0 \\
 \tt 1110101001001000 & \tt EA48 & \tt 1111110010100000 & \tt FCA0 \\
 \tt 1110110011001000 & \tt ECC8 & \tt 1111110010001000 & \tt FC88 \\
 \tt 1111100011001000 & \tt F8C8 & \tt 1111101010100000 & \tt FAA0 \\
 \tt 1110101011100000 & \tt EAE0 & \tt 1111101010001000 & \tt FA88 \\
 \tt 1110110011100000 & \tt ECE0 & \tt 1110111010100000 & \tt EEA0 \\
\hline
\end{tabular}
\caption{The $20$ (out of $2^{16}=65536$) rules for $K=4$ random
  Boolean networks that successfully solve the density classification
  task for all initial configurations.}
\label{tab:k4rules}
\end{center}
\end{table}

\subsubsection{$K=5$: Many Polynomials}
By the same procedure we find the values of $a_j$ when $K=5$:
\[a_0 = 0,\quad a_1 = 0,\quad  a_2 \in [0,5], \quad a_3 = 10-a_2,\quad
a_4 = 5\quad {\rm and}\quad a_5 = 1.\]

Since $a_2$ is the only free parameter, $6$ polynomials satisfy
conditions C1 and C2, each corresponding to a value of $a_2$. The {\sl
majority} rule is the only possible rule when $a_2=0$, whereas in the
other cases, each polynomial represents many rules. For example, when
$a_2=3$, ${10 \choose 3} {10 \choose 7} = 14,400$ equivalent rules
exist.

As one can see, representing the rules by means of a polynomial bears
the advantage that rules with similar dynamics are grouped together by
Equation \ref{eq:poly2}. They may thus more easily be handled and
analyzed on a meta-level.

\subsection{Rules for the Synchronization Task}
\label{sec:sync_rules}
The procedure to find the rules for the synchronization task is
similar to the density classification task. Again, we are looking for
maps $p_1^{t+1} = Q_K(p_1^t)$ similar to those shown on Figure
\ref{fig:firefly_poly}.  More formally, the constraints we have to
impose on $Q_K(x)$ are firstly
\[Q_K(0) = 1\quad{\rm and}\quad Q_K(1) = 0,\]
which will force the network to oscillate between all $0$s and all
$1$s on consecutive time steps. And secondly, in order to make sure
that the network's state converges towards $n_1^t = 0$ or $n_1^t = 1$,
$Q_K''(x)$ needs to be monotone and $Q_K''(x) \neq 0$ for all $x \in
[0,1]$.

Two functions satisfying these constraints are for example:
\[Q_K(x) = (1-x)^K\quad{\rm and }\quad Q_K(x) = \sum_{i=0}^{K-1}
x^i\,(1-x)^{K-i}.\]
These functions are symmetrical and therefore perform equally well. We
will call them $\gamma_K$-rules.  Their behavior is very simple: the
lookup-table associated with this rule outputs a $1$ (respectively
$0$) if and only if all input bits are $0$ (respectively $1$).
Examples are rule $\gamma_3=1$ and rule $\gamma_3=128$ for $K=3$
networks. Both rules are basically fully equivalent and solve the
synchronization task in the same way.

\section{Experiments and Results}
\label{sec:experiments}
In this section we will examine the performance of RBNs as a function
of the number of nodes $N$ and the number of connections per node $K$.
Since the number of initial configurations exponentially grows with
$N$, testing them all becomes rapidly computationally intractable.
Algorithm \ref{alg:sampling} illustrates the procedure we used for all
experiments in this section. Since the nodes are randomly connected,
we average the simulation results on a certain number of randomly
generated networks. In addition, each initial configuration is
generated according to a biased distribution, i.e., the number of $1$s
it contains is randomly chosen from a uniform distribution and the
$1$s are then assigned to randomly chosen nodes.  This will force the
network to be tested uniformly on all possible values of $p_1^0$ and
not only around $p_1^0 = 1/2$, which is the case if each configuration
is chosen according to an unbiased distribution, i.e., each node has a
uniform probability to be in state $1$ or $0$.
\begin{algorithm}
\begin{algorithmic}
  \STATE $N_{\rm min} = 9$ (minimal number of nodes)
  \STATE $N_{\rm max} = 201$ (maximal number of nodes)
  \STATE $\texttt{rep} = 5000$ (number of initial configurations)
  \STATE $\texttt{nnet} = 500$ (number of networks)
  \FOR{$n \in [N_{\rm min},N_{\rm max}]$}
  \FOR{$i \in [1,\texttt{nnet}]$}
  \STATE Generate a random network composed of $n$ nodes.
  \FOR{$r \in [1,\texttt{rep}]$}
  \STATE Generate an initial configuration.
  \STATE Test the configuration on the network.
  \ENDFOR
  \STATE Average the results over $\texttt{rep}$.
  \ENDFOR
  \STATE Average the results over $\texttt{nnet}$.
  \ENDFOR
\end{algorithmic}
\caption{Performance Evaluation Algorithm}
\label{alg:sampling}
\end{algorithm}
Finally, in order to compare RBNs with CAs, we shall only test
networks with an odd number of nodes. This avoids the unstable fixed
point in $p_1^0 = 1/2$, which will be misclassified most of the time
anyway, since, as explained in Section \ref{sec:density_task}, a small
perturbation, such as changing the state of one node, might be
sufficient to drive the system towards the wrong solution.

\subsection{The Density Classification Task}
\label{sec:perfs}
Figure \ref{fig:density_perf} shows the network performance for $K \in
\{3,4,5,6,7,8,9\}$. All networks were uniform and simply used the
majority rule, which, as shown in Section \ref{sec:density_rules}, is
the only common solution when $K \geqslant 3$.

\iclfig{scale=0.8}{density_perf.0}{Density classification task
performance as a function of $N$ and $K$.  (1) solid line: $K=3$,
dashed line: $K=4$, (2) bottom-up, $K = 5,7,9$, (3) bottom-up,
$K=8,6$.}{fig:density_perf}

The results show that the performance increases with increasing $N$.
We will later show that it tends to $100\%$ when $N \to +\infty$.
Moreover, for odd values of $K$, RBNs perform better than for even
values, and the larger $K$, the better the performance. This finding
confirms that the majority rule works better without ambiguous
situations, i.e., when a node does not receive an equal number of $1$s
and $0$s.

Figure \ref{fig:density_len} shows the average number of iterations a
RBN requires to converge to a solution. Networks that need a small
number of iterations to solve the density classification task are said
to perform better. One can see that the number of iterations required
increases with the network's size $N$.  This intuitively makes sense
as a larger network needs more time to solve the task because the
information must be propagated among the nodes.  However, it is worth
mentioning that the number of iterations required does not increase
linearly with $N$, as one might expect, but rather follows a
logarithmic-like law. This can be explained by the fact that random
networks allow for long-distance connections, which help to propagate
information faster.

\iclfig{scale=0.8}{density_len.0}{Average number of iterations
required to solve the density classification task as a function of $N$
and $K$.  (1) solid line: $K=3$, dashed line: $K=4$, (2) bottom-up,
$K=9,7$, (3) bottom-up, $K=5,8,6$.}{fig:density_len}

The performance increase with increasing $N$ can be explained by means
of Equation \ref{eq:poly}: although the polynomials are continuous
functions, we evaluate them at discrete time steps, i.e., $p_1^t =
n_1^t/N$ can only take a finite number of discrete values.  For
example, let us suppose a network with $N=10$ and $K=3$. From a
theoretical point of view, if $p_1^t = 6/10$, then $p_1^{t+1} =
0.648$, however, since $p_1^{t+1} \in \{i/N\,|\,i\in[0,N]\}$ the exact
value taken by $p_1^{t+1}$ will then depend on the actual network
wiring.  Performance will therefore get better with increasing $N$ and
tends to perfection when $N \to +\infty$.  In order to further
illustrate this, we generated Figures \ref{fig:density-n53-k3},
\ref{fig:density-n53-k5}, \ref{fig:density-n501-k3}, and
\ref{fig:density-n501-k5}, that include both the theoretical and the
experimental results.

From these four figures we can see that the bigger $N$, the better the
theoretical curves are matched. As expected, we observe large
variances for small values of $N$, which implies that the network may
converge toward the wrong solution near the fixed point $p_1^* = 1/2$,
since $p_1^{t+1}$ may cross the identity function $Q_K(x) = x$.  This
explains why the density classification task is solved less
successfully for small $N$.

\iclfig{scale=0.8}{density-n53-k3.0}{Density classification task map
  $p_1^{t+1} = Q_K(p_1^t)$ for $N=53$ and $K=3$. Mean (diamond)
  simulated and theoretical curve (dashed line). The identity function
  is also shown (dashed line). The bars indicate the maximum and
  minimum values observed.}{fig:density-n53-k3}

\iclfig{scale=0.8}{density-n53-k5.0}{Density classification task map
  $p_1^{t+1} = Q_K(p_1^t)$ for $N=53$ and $K=5$. Mean (diamond)
  simulated values and theoretical curve (dashed line). The identity
  function is also shown (dashed line). The bars indicate the maximum
  and minimum values observed.}{fig:density-n53-k5}

\iclfig{scale=0.8}{density-n501-k3.0}{Density classification task map
  $p_1^{t+1} = Q_K(p_1^t)$ for $N=501$ and $K=3$. Mean (diamond)
  simulated values and theoretical curve (dashed line). The identity
  function is also shown (dashed line). The bars indicate the maximum
  and minimum values observed.}{fig:density-n501-k3}

\iclfig{scale=0.8}{density-n501-k5.0}{Density classification task map
  $p_1^{t+1} = Q_K(p_1^t)$ for $N=501$ and $K=5$. Mean (diamond)
  simulated values and theoretical curve (dashed line). The identity
  function is also shown (dashed line). The bars indicate the maximum
  and minimum values observed.}{fig:density-n501-k5}

Moreover, the performance for odd values of $K$ is better because the
derivative at the fixed point $p_1^* = 1/2$ tends to be infinite when
$K \to N$.  Errors in that point are then less likely to make
$p_1^{t+1}$ cross the identity function $Q_K(x)=x$, and therefore to
drive the network towards the wrong solution. This can easily be seen
if Figures \ref{fig:density-n501-k3} and \ref{fig:density-n501-k5}:
for $K=3$, there are at least four points in the neighborhood of
$p_1^* = 1/2$ with an extrema that crosses the identity function,
while there are only two for $K=5$. Finally, note that the density
classification task is trivially solved in one time step by the
majority rule when $K=N$.

\subsection{The Synchronization Task}
In order to evaluate the performance of RBNs on the synchronization
task, we used the following set of parameters (see Algorithm
\ref{alg:sampling}): $N_{\rm min} = 10$, $N_{\rm max} = 200$,
$\texttt{rep} = 10,000$ and $\texttt{nnet} = 200$. However, the
initial configurations were still chosen according to a biased
distribution as defined above.

Figure \ref{fig:firefly_perf} shows the performance of the
synchronization task, whereas Figure \ref{fig:firefly_len} shows the
number of iterations necessary to synchronize the automaton. The
network was uniform and all nodes used the $\gamma_K$-rule as
described in Section \ref{sec:sync_rules}.  One can see that,
similarly to the density classification task, performance increases
with increasing $N$ and $K$ and that it tends to perfection (i.e.,
$100\%$) when $N \to +\infty$ or $K \to N$.  However, perfect results
are already obtained for $N \approx 150$ and $K > 3$. Note that if
$K=N$, each node possesses a full view on the entire network state and
the task therefore becomes trivial.

\iclfig{scale=0.8}{firefly_perf.0}{Synchronization task performance as
a function of $N$ and $K$. Bottom-up, $K =
2,3,4,5,6,7,8,9$.}{fig:firefly_perf}

\iclfig{scale=0.8}{firefly_len.0}{Average number of iterations
required before synchronization occurs as a function of $N$ and
$K$. Top-down, $K=2,3,4,5,6,7,8,9$.}{fig:firefly_len}

\subsection{RBNs versus CAs}
\label{sec:rbns_vs_cas}
The performance of CAs for both the density classification and the
synchronization task were investigated in various publications (see
Section \ref{sec:density_task} and \ref{sec:sync_task}). We shall
compare these results in the present section to those obtained with
RBNs and will especially focus on how the information is processed in
CAs and RBNs.

In \cite{mitchelletal96}, Mitchell et al. presented four different CA
rules for the density classification task, namely $\phi_{\rm GKL}$,
$\phi_{\rm maj}$, $\phi_{\rm exp}$, and $\phi_{\rm par}$, which all
used a CA neighborhood of radius $r=3$, i.e., each cell is connected
to its $2r$ nearest neighbors and to itself. $\phi_{\rm maj}$ is the
majority rule, $\phi_{\rm exp}$ and $\phi_{\rm par}$ were generated by
means of an evolutionary algorithm, and $\phi_{\rm GKL}$ is a
hand-designed rule derived from Gacs, Kurdyumov and Levin's work
\cite{gonzagaetal92,gacsetal78}, which is defined as following:
\[s_i^{t+1} = \left\{\begin{array}{cl}
  \textrm{majority}\;(s_i^t, s_{i-1}^t, s_{i-3}^t) & \textrm{if $s_i^t
  = 0$} \\ 
  \textrm{majority}\;(s_i^t, s_{i+1}^t, s_{i+3}^t) & \textrm{if $s_i^t
  = 1$} 
\end{array}\right..\]
$\phi_{\rm exp}$\footnote{Rule table for a $r=3$ CA: {\tt
    0505408305c90101200b0efb94c7cff7}.} uses of a block expansion
\cite{mitchelletal96}, whereas $\phi_{\rm par}$\footnote{Rule table
  for a $r=3$ CA: {\tt 0504058705000f77037755837bffb77f}.} uses a
``particle-based'' strategy \cite{mitchelletal96}. Furthermore, we
considered the hand-written Das-rule \cite{dasetal94}, the
Andre-Bennett-Koza (ABK) rule \cite{andre96}, which was found by means
of a genetic algorithm, as well as two others rules generated by means
of co-evolution \cite{juille98}. Those rules will be compared to three
RBN rules, namely $\vartheta_3, \vartheta_5, \vartheta_7$, where
$\vartheta_i$ is the majority rule for $K=i$. In order to compare
two-states CAs with RBNs, the most natural way is to chose a RBN with
a $K=2r + 1$ neighborhood, however, as we will see, even RBNs with $K
< 2r + 1$ perform better than CAs with a $r$-neighborhood.

In \cite{mitchelletal96}, the performance $\mathcal{P}_{N,10^4}$ of a
network composed of $N$ nodes is defined as the fraction of correct
classifications made over $10^4$ initial configurations that are
randomly selected from an unbiased distribution, i.e., each bit in the
initial configuration is randomly chosen. The expected percentage of
$1$s in a configuration if therefore $1/2$ (see Section
\ref{sec:experiments}).  For our RBNs, $\mathcal{P}_{N,10^4}$ is
averaged over $200$ randomly generated networks.

Table \ref{tab:dens_results} summarizes the performance of the density
classification task for the different rules. One can see that the
$\vartheta$-rules outperform the $\phi$-rules since $\vartheta_3$ is
as good as the best rule found by the evolutionary algorithm for
$N=149$, and it performs better than $\phi_{\rm GKL}$ for $N=599$ or
$N=999$.  Furthermore, as mentioned in \cite{dasetal94}: ``[t]he
performance of these rules decreased dramatically for larger $N$
[\ldots],'' which is not the case for RBNs, on the contrary, they
perform better with increasing $N$ (see Figure
\ref{fig:density_perf}).

\begin{table}
\begin{center}
\begin{tabular}{|c|c|c|c|c|c|}
\hline
Type & Rule & $\mathcal{P}_{149,10^4}$ & $\mathcal{P}_{599,10^4}$ & $\mathcal{P}_{999,10^4}$ & Reference \\
\hline
CA & $\phi_{\rm maj}$  & 0.000       & 0.000       & 0.000       & \cite{mitchelletal96} \\
CA & $\phi_{\rm exp}$  & 0.652       & 0.515       & 0.503       & \cite{mitchelletal96}\\
CA & $\phi_{\rm par}$  & 0.769       & 0.725       & 0.714       & \cite{mitchelletal96}\\
CA & $\phi_{\rm GKL}$  & 0.816       & 0.766       & 0.757       & \cite{mitchelletal96}\\
CA & Das rule          & 0.823       & 0.778       & 0.764       & \cite{dasetal94} \\
CA & ABK rule          & 0.824       & 0.764       & 0.730       & \cite{andre96} \\
CA & Co-evolution (1)  & 0.851       & 0.810       & 0.795       & \cite{juille98} \\
CA & Co-evolution (2)  & {\bf 0.860} & 0.802       & 0.786       & \cite{juille98} \\
\hline
RBN & $\vartheta_3$    & 0.766       & 0.771       & 0.769       & Mesot and Teuscher \\
RBN & $\vartheta_5$    & 0.823       & 0.825       & 0.820       & Mesot and Teuscher \\
RBN & $\vartheta_7$    & 0.850       & {\bf 0.848} & {\bf 0.852} & Mesot and Teuscher \\
\hline
\end{tabular}
\caption{Density classification performance of CAs and RBNs for
  different network sizes and different rules. The best result in each
  column is shown in bold.}
\label{tab:dens_results}
\end{center}
\end{table}

In 1995, Das et al. \cite{dasetal95} described the results obtained by
four evolved $\phi$-rules on the synchronization task. Those rules
where used on CAs with radius $r=3$ and were evaluated by the same
procedure as Mitchell et al. \cite{mitchelletal96} used for the
density classification task. Table \ref{tab:sync} summarizes the
performance of $\phi$- and $\gamma$-rules for different network sizes.
One can see that all $\gamma_K$-rules as well as the $\phi_{\rm
  sync}$-rule are able to perfectly solve the synchronization task. We
may wonder why a RBN of $149$ nodes does not make any error since we
have seen in Figure \ref{fig:firefly_perf} that a network of $200$
nodes did not perform perfectly? The difference lies in the fact that
performance is essentially evaluated on configurations where the
percentage of $1$s is around $1/2$.

\begin{table}
\begin{center}
\begin{tabular}{|c|c|c|c|c|c|}
\hline
Automaton & Rule & $\mathcal{P}_{149,10^4}$ & $\mathcal{P}_{599,10^4}$ & $\mathcal{P}_{999,10^4}$ & Reference \\
\hline
CA & $\phi_1$ & 0.00 & 0.00 & 0.00 & \cite{dasetal95}\\
CA & $\phi_2$ & 0.33 & 0.07 & 0.03 & \cite{dasetal95}\\
CA & $\phi_3$ & 0.57 & 0.33 & 0.27 & \cite{dasetal95}\\
CA & $\phi_{\rm sync}$ & 1.00 & 1.00 & 1.00 & \cite{dasetal95}\\
\hline
RBN & $\gamma_3$ & 1.00 & 1.00 & 1.00 & Mesot and Teuscher\\
RBN & $\gamma_5$ & 1.00 & 1.00 & 1.00 & Mesot and Teuscher\\
RBN & $\gamma_7$ & 1.00 & 1.00 & 1.00 & Mesot and Teuscher\\
\hline
\end{tabular}
\end{center}
\caption{Synchronization performance of CAs and RBNs for different 
network sizes and different rules.}
\label{tab:sync}
\end{table}

\subsection{An Entropy-Based Performance Measure}
\label{sec:entropy_measure}
In order to measure a rule's capability to solve the density
classification or the synchronization task, we will introduce the
following measure derived from Wuensche's {\sl input entropy}
\cite{wuensche99}:
\beq
H(W_1^t\,|\,P_1^t) = \sum_{i=0}^N H\left(W_1^t\,|\,P_1^t = i/N\right) P\left(P_1^t = i/N\right),
\label{eq:measure}
\eeq
where $H(\cdot)$ is the Shannon entropy \cite{shannon48}. This measure
simply computes the conditional entropy of the number of $1$s in a
node's input, knowing the state of the network at time $t$, i.e., the
percentage of $1$s, $P_1$. Contrary to the measure used by Wuensche,
which considers each possible node input, our measure only takes into
account the distribution of $W_1^t$, i.e., the number of $1$s in an
input. This means that two node input configurations that contain the
same number of $1$s are considered to be equivalent. Our measure not
only allows to qualitatively compare the CA and RBN behavior, but it
also gives indications on how difficult it is for a rule to solve a
given task and what the average number of required times steps is. For
example, if we obtain $H(W_1^t \,|\, P_1^t) = 0$ after a certain
amount of time, we know that the network has converged to an all $1$s
or all $0$s state. This will therefore provide us with an estimation
of how fast a RBN may solve a task.

We have seen in Section \ref{sec:comp_and_inf} that $W_1^t$ follows a
binomial law $\mathcal{B}(K, p_1^t)$, where $p_1^t$ is implicitly
defined by means of the map $p_1^t = Q_K(p_1^{t-1})$ and by an initial
distribution function for $P_1^0$. For example, Figure
\ref{fig:bin_th} shows the variation of the input entropy if we use
the map defined by Equation \ref{eq:density_k3} (Section
\ref{sec:density_task}) under the assumption that the initial
configurations are randomly chosen according to a biased distribution,
i.e., $P(P_1^0) = \mathcal{U}(0,1)$\footnote{$\mathcal{U}(a,b)$ is the
  uniform distribution on the interval $[a,b]$.}. As can be seen, the
input entropy quickly decreases towards zero. This means that---at
least theoretically---it takes on average less than $15$ time steps to
settle down in a fixed point when $K=3$ and the majority rule is
used. However, as shown in Figure \ref{fig:density_len}, the actual
number of steps required is much lower (the average lies around $4.5$
steps).  Figures \ref{fig:entropy_bias} and \ref{fig:entropy_unbias}
show the variation of the conditional input entropy for the RBN rule
$\vartheta_7$ as well as for the CA rules $\phi_{\rm maj}$, $\phi_{\rm
  exp}$, $\phi_{\rm par}$, and $\phi_{\rm GKL}$ ($N=149$, initial
configurations selected using a biased or unbiased distribution
respectively, see Section \ref{sec:rbns_vs_cas}).

\iclfig{scale=0.8}{bin_th.0}{Theoretical variation of the conditional
input entropy, $H(W_1^t\,|\,P_1^t)$, when a RBN solves the density
classification task.}{fig:bin_th}

\iclfig{scale=0.8}{entropy_bias.0}{Variation of the conditional input
  entropy of different rules for the density classification
  task. Initial configurations are chosen from a biased distribution
  and $N=149$.  (1) solid line: $\phi_{\rm GKL}$, dashed line:
  $\phi_{\rm par}$.}  {fig:entropy_bias}

\iclfig{scale=0.8}{entropy_unbias.0}{Variation of the conditional
  input entropy of different rules for the density classification
  task. Initial configurations are chosen from an unbiased
  distribution. $N=149$.  (1) solid line: $\phi_{\rm GKL}$, dashed
  line: $\phi_{\rm par}$.}  {fig:entropy_unbias}

The results suggest the following comments:
\begin{itemize}
\item $\vartheta_7$ exactly follows the curve as predicted by our
  theory.
\item All plots start slightly below a value of $2$. This value is
  exactly the same as for a binomial law, which means that at time
  $t=0$, no difference exists between RBNs and CAs.
\item When $t>0$, the behaviors of the $\phi$-rules are fairly
  different from the $\vartheta$-rules. As seen in Section
  \ref{sec:entropy_measure}, RBNs follow a simple curve that
  corresponds to a binomial law. CAs, on the other hand, need more
  complex behaviors to achieve the same result.
\item RBNs with $\vartheta_7$-rule have a very simple, straightforward
  behavior compared to CAs. After a very small number of iterations,
  the input entropy is already less than one bit, which means that
  most of the nodes basically see the same input.
\item The $\phi_{\rm maj}$-rule is very similar to the
  $\vartheta_7$-rule for unbiased distributions. However, because of
  the CA's connection topology, frozen blocks show up and prevent the
  automata from converging to a zero input entropy.
\item The behavior of the best evolved $\phi_{\rm par}$-rule and of
  the hand-designed $\phi_{\rm GKL}$-rule are very similar. Note that
  $\phi_{\rm GKL}$ has intentionally been designed to break the
  inherent symmetry of CAs and thus requires more time to converge.
\item The behavior of the $\phi_{\rm exp}$-rule is very similar to
  that of the $\vartheta_7$-rule. However, Figure
  \ref{fig:entropy_unbias} shows that an increase in complexity is
  required around $H(W_1^t\,|\,P_1^t) = 1$ to overcome the emergence
  of frozen blocks.  $\phi_{\rm par}$ and $\phi_{\rm GKL}$, on the
  other hand, increase the complexity of their behavior from the very
  beginning in order to somehow compensate for the regularity of the
  interconnection topology.
\end{itemize}

Similarly, Figures \ref{fig:ff_h_bias} and \ref{fig:ff_h_unbias} show
the variation of the conditional input entropy for the synchronization
task. The $\gamma_7$-rule as well as the three CA rules $\phi_2$,
$\phi_3$ and $\phi_{\rm sync}$ were used, $N=149$, and initial
configurations were selected using a biased and unbiased distribution
respectively (see Section \ref{sec:rbns_vs_cas}). The results suggest
the following comments:
\begin{itemize}
\item $\gamma_7$ follows almost exactly the curve as
  predicted by our theory. A small difference, however, exists when
  the unbiased distribution is used (Figure \ref{fig:ff_h_unbias}).
\item In the unbiased case, the system converges after less than three
  iterations, which is very rapid, given that the synchronization is
  always perfect in that case (see Table \ref{tab:sync}).
\item The input entropy is not fully monotonous (Figure
  \ref{fig:ff_h_unbias}).  This can be explained by looking at the
  evolution of the network's state during the first three iterations:
  after the first iteration, most of the nodes are in state $0$ since
  $\gamma_7$ produces a $1$ only if the input contains $0$s only. The
  input entropy is thus very low, however, at the second iteration,
  all nodes which are connected to a $1$ will be in state $0$ and all
  others in state $1$ since they are connected to $0$s only. Thus, a
  single $1$ in the network after the first iteration can generate
  more than one $0$ at the next step and therefore increases the input
  entropy.
\item The non-convergence of $\phi_2$ and $\phi_3$ after $300$
  iterations indicates the existence of frozen-blocks, which makes the
  problem much harder to solve for these rules.
\end{itemize}

\iclfig{scale=0.8}{ff_h_bias.0}{Variation of the conditional input
  entropy of different rules for the synchronization task. Initial
  configurations are chosen from a biased distribution,
  $N=149$.}{fig:ff_h_bias}

\iclfig{scale=0.8}{ff_h_unbias.0}{Variation of the conditional input
  entropy of different rules for the synchronization task. Initial
  configurations are chosen from an unbiased distribution, $N=149$.
  Subgraph: solid line = simulated $\gamma_7$, dashed line =
  theoretical $\gamma_7$.}  {fig:ff_h_unbias}

\section{Non-Standard Architectures and Topologies}
\label{sec:non-standard}
In order to improve the computational power of CAs, various people
proposed to introduce non-uniformity among the rules. Although this
approach sacrifices one of the main advantages of CAs, namely, to have
simple and identical processing elements, it allows to find better
solutions \cite{sipper94:alife,sipper96:physicad,sipper99:tcs}. These
solutions often rely on a particular structure where the nodes' rules
must be carefully chosen. With RBNs, however, this is not the case: as
we have seen in Section \ref{sec:rules}, when $K=5$ for example, a
single map can represent $14,000$ rules, which are strictly
equivalent. We can thus easily build non-uniform RBNs by randomly
choosing one rule for each node instead of using the same rule in all
nodes.

CAs and RBNs can be considered as two extreme classes of networks that
are not really observed in Nature. From the point of view of the
interconnection topology, CAs are simply too regular, whereas random
networks require an equal probability for each connection to be
established among all nodes, which is in most cases subjected to
physical limitations since long-distance connections are more costly
than short-distance connections. Many recent studies have instead
confirmed that an important number of networks observed in Nature
belong to a class called {\sl small-world} networks (see for example
\cite{watts98,strogatz01:_explor,barabasi99:_emerg_scalin,albert02}).
These networks may be considered as an intermediate class between CAs
and RBNs, since, if we gradually transform a CA into a RBN in a
certain manner, we can obtain small-world networks.

Different types of small-world networks exist \cite{amaral00}, here we
shall however only consider the networks of Watts and Strogatz as
described in \cite{watts98}. In their paper they state for the density
classification task that ``[\ldots] a simple {\sl majority-rule}
running on a small-world graph can outperform all known human and
genetic algorithm-generated rules running on a ring lattice.''
However, they do not give the amount of randomness a network should
contain in order to outperform CAs on the density task. Moreover,
since we also studied the synchronization task, it would be
interesting to see if small-world networks using the $\gamma$-rule
perform better than CAs on this task as well. We performed several
experiments with networks built in the following way:

\begin{enumerate}
  \item Start with a ring lattice, which might be considered as a
  special one-dimensional CA with a $r=1$ neighborhood, but without
  self-connections.
\item Rewire the connections with probability $\rho$ as described in
  \cite{watts98}.
\end{enumerate}

Therefore, if $\rho = 0$, the initial automaton is left unchanged, if
$\rho=1$, it is transformed into an automaton with a random topology.
Note that, although the connectivity $K$ was $2$ for each node at the
beginning, each node can potentially have a different connectivity at
the end. The average of $K=2$ connections per node is, however, not
affected by this random rewiring algorithm.

Figures \ref{fig:sw-perf} and \ref{fig:sw-len} show the percentage of
success and the number of iterations required for solving the density
classification and the synchronization task as a function of $\rho$,
i.e., the amount of randomness in the network. The simulations were
performed for a network size of $N=149$ nodes, the rules used were the
majority rule for the density classification and the $\gamma$-rule for
the synchronization task.

The results can be summarized as following: For both tasks, the
percentage of success reaches its maximal value around $\rho=0.5$ (see
Figure \ref{fig:sw-perf}). Whereas the number of iterations for the
synchronization task reaches its maximum around $\rho=0.3$ (Figure
\ref{fig:sw-len}) already, the density classification task requires
significantly less iterations on a random topology. The reason for
this can be found in the corresponding maps (Figures
\ref{fig:density_poly} and \ref{fig:firefly_poly}): the
synchronization map does not possess stable fixed points, which makes
it less sensitive to additional connections between the already
existing clusters.

\iclfig{scale=0.8}{sw-perf.0}{Percentage of success of small-world
  networks for the density classification (solid line) and the
  synchronization task (dashed line) as a function of the amount of
  randomness $\rho$. $N=149$, majority rule (density) and
  $\gamma$-rule (synchronization).}  {fig:sw-perf}

\iclfig{scale=0.8}{sw-len.0}{Number of iterations required by
  small-world networks for solving the density classification (solid
  line) and the synchronization task (dashed line) as a function of
  the amount of randomness $\rho$. $N=149$, majority rule (density)
  and $\gamma$-rule (synchronization). }{fig:sw-len}

We can conclude that a relatively small amount of randomness already
greatly helps to improve performance for solving our two tasks.
However, especially for the density classification task, a randomly
interconnected network helps to significantly improve
convergence. Despite the extreme simplicity of our random Boolean
networks used, the results fit into the global picture of recent work
on complex network synchronization using coupled oscillators (see for
example \cite{motter05,nishikawa03}), although more works is certainly
needed in that area.

\section{Conclusions}
\label{sec:conclusions}
In this paper we have first derived a recursive equation which gives
the probability of a node to be in state $1$ at time $t+1$ for random
Boolean networks. Based on that equation, we were then able to
analytically find the local node rules from the global network
behavior for our two comparative tasks.  Our main findings are the
following:

\begin{enumerate}
\item No $K=2$ random Boolean network can perfectly, i.e., for all
  initial configurations, solve the density classification task. The
  majority rule is the only perfect rule for $K=3$ networks, whereas
  many rules exist for $K>3$ networks.
  
\item The two symmetrical $\gamma_K$-rules are the best rules for
  solving the synchronization task.
  
\item Random Boolean networks outperform cellular automata on both the
  density classification and on the synchronization task.
  
\item Random Boolean networks also outperform networks with a
  small-world topology on the density classification task, but their
  performance is equivalent for the synchronization task.
  
\item The rules obtained for both our tasks are highly scalable, i.e.,
  the larger the network, the better they perform. This is not the
  case for cellular automata.
\end{enumerate}

The reason why random Boolean networks perform better on both the
synchronization and on the density classification tasks is the
following: due to the random wiring, each node of a random Boolean
network has---from a statistical point of view---an unbiased view on
the automata's global state since the distribution of $1$s in the node
inputs of each node is the same as in the global network state. On the
other hand, cellular automata have a biased view because of their
local neighborhood, which makes it more difficult for them to solve
global tasks.

Another advantage of random Boolean networks over CAs is that they
allow the network to be rewired at any time, without affecting the
global performance. This property is especially interesting if
interconnections are likely to fail. One would then only have to
re-establish a new connection to a randomly chosen alternative node.
This property could be very interesting for large scale distributed
systems. Although not explicitly tested in this work, the existence
of many equivalent rules for certain values of $K$ tend to show that
RBNs may exhibit great robustness to node and interconnection
failures.

One might of course ask whether the current approach might be applied
to similar global tasks and whether it might be generalized to
non-uniform random Boolean networks, alternative topologies, or
asynchronously updating nodes. Our current work concentrates exactly
on these questions. The goal is to extend the theory to asynchronously
updating and non-uniform random Boolean networks and to other, more
useful tasks.  Asynchrony not only allows to remove the sole global
clock signal necessary to synchronously update the cells, which often
represents a constraint for hardware realizations, but also yields in
further potentially interesting properties.  Asynchronous automata
attracted much interest in the past few years, although it has been
proved independently by Capcarr\`ere \cite{capcarrere02_thesis} and
Nehaniv \cite{nehaniv2003:alife} that any $n$-state synchronous
automata can be emulated by a particular $3n^2$-state asynchronous
automata. In their 1984 experimental study, Ingerson and Buvel
\cite{IngandBuw:84} already explored the question of how much of the
interesting behavior of cellular automata comes from the synchronous
modeling. They concluded---based on the visual appearance of the
evolution patterns over time---that the synchronous assumption is not
essential to the study of cellular automata and that certain
irrelevant structures may appear from the synchronous update of the
cells.  However, one of the main arguments against purely synchronous
automata as a tool for modeling has always been the lack of biological
plausibility \cite{stark:biosystems2000}, although a purely
asynchronous behavior is certainly not biologically plausible
either. But there are several other issues of interest in asynchronous
models: Bersini and Detours \cite{bersinietal94} suggested that
asynchrony might induce stability, but also forces evolution to find
more robust solutions, as Rohlfshagen and Di Paolo have shown for the
case of rhythmic asynchronous random Boolean networks
\cite{rohlfshagen04}.

A further topic of interest is also the extension of the theory to
$S$-state random networks, which are likely to have interesting
properties and would represent a generalization of $S$-state CAs.  The
formalization of asynchronously updating and non-uniform random
Boolean or $S$-state networks and the ability to derive their local
rules for a larger class of global tasks would certainly represent a
major step in that field of research, where the local node rules are
commonly either hand-designed or evolved so far.

\subsection*{Acknowledgments}
The authors are grateful to Jonas Buchli for his helpful comments and
discussions. This work was supported in part by the Swiss National
Science Foundation under grant PBEL2--104420.


\end{document}